\documentclass[1p]{elsarticle}

\usepackage{amssymb}
\usepackage{amsfonts}
\usepackage{amstext}
\usepackage{graphicx}
\usepackage{amscd}
\usepackage{amsmath}
\usepackage{mathrsfs}
\usepackage{stmaryrd}

\newcommand{\tr}{\mathrm{tr}}
\newcommand{\RE}{\Re\mathrm{e}}
\newcommand{\IM}{\Im\mathrm{m}}
\newcommand{\CO}{\mathfrak{C}\mathrm{o}}
\newcommand{\iquat}{\mathbf{i}}
\newcommand{\jquat}{\mathbf{j}}
\newcommand{\kquat}{\mathbf{k}}
\newcommand{\ihbar}{\iquat \hbar}

\begin{document}
\begin{frontmatter}

\title{Competition between decoherence and purification: quaternionic representation and quaternionic fractals}

\author{David Viennot}
\address{Institut UTINAM (CNRS UMR 6213, Universit\'e de Bourgogne-Franche-Comt\'e, Observatoire de Besan\c con), 41bis Avenue de l'Observatoire, BP1615, 25010 Besan\c con cedex, France.}

\begin{abstract}
We consider the competition between decoherence processes and an iterated quantum purification protocol. We show that this competition can be modelled by a nonlinear map onto the quaternion space. This nonlinear map has complicated behaviours, inducing a fractal border between the area of the quantum states dominated by the effects of the purification and the area of the quantum states dominated by the effects of the decoherence. The states on the border are unstable. The embedding in a 3D space of this border is like a quaternionic Julia set or a Mandelbulb with a fractal inner structure.
\end{abstract}

\end{frontmatter}


\begin{quotation}
 Qubits are the resource of the quantum information as bits for the classical information, and are the main subject for future technologies as quantum computers. In contrast with bits, qubits exhibit states which are impossible at a classical level as Schr\"odinger cat states (the qubit is in a superposition of 0 and 1). These purely quantum properties are the resource to drastically increase the performance of the computing. But the noises of the environment generate a physical process called \textit{decoherence} which suppresses the purely quantum properties. There is a protocol, called \textit{purification}, which permits to restore the quantum behaviour. The result of the competition between a permanent decoherence process and a repeated purification protocol is not simple because this generates a chaotic process. We show that this one is a generalisation of the famous Julia map (which generates the famous fractals known as the Julia and the Mandelbrot sets). More precisely, in place of a map of the complex plane, the decoherence-purification competition map is a map of the quaternionic space (so-called Hamilton's number set, which are numbers which do not commute, i.e. $zw \not= wz$ with $z$ and $w$ two quaternions). The decoherence-purification map generates 3D fractal sets similar to a Mandelbulb (3D generalisation of a Mandelbrot set) with a fractal inner structure.
\end{quotation}

\section{Introduction}
Decoherence is a physical process consisting to the lost of the quantum properties due to the environment effects. Under decoherence, the purity decreases (this one measures the pure quantum behaviour of a state, see \cite{Bengtsson}). Decoherence can result from entanglement between the quantum system and its environment \cite{Breuer}, from chaotic or stochastic noises induced by the environment \cite{Viennot} or from thermal fields emitted by the environment \cite{Breuer}. Some researches hope to use the quantum laws for practical applications, as quantum teleportation \cite{Heiss}, quantum computing \cite{Heiss} and quantum control \cite{Brif}. For these goals, the decoherence processes are hampers ruining the attempts to reach the desired targets. Rather than trying to narrow the decoherence processes (as in usual strategies), we could try to fight them by using a purification protocol. Such a one, as for example in \cite{Benchmann} for a qubit (quantum bit), consists to manipulate the quantum system in order to increase the purity of its state. Formally, the purification is a nonlinear map of the state space, which is physically realised by entanglement, quantum measurement and post-selection (see \cite{Benchmann,Terno} for details). By repeating a purification protocol, we want to fight against the decoherence. The question is then: Is the purification or the decoherence which wins the competition? We can imagine that the answer depends on the initial mixed state $\rho$. A second question is then: what is the behaviour of the states at the border between the area dominated by the decoherence and the area dominated by the purification? The nonlinearity of the purification protocol induces some complicated behaviours. As shown in \cite{Kiss1,Kiss2,Kiss3}, if we repeat the purification protocol onto pure states, some of them are stable (the pure state orbit reaches cyclic points) but some other states are unstable. The border between the two behaviours is a fractal set.\\
A simple map of the complex plane inducing a complicated behaviour is for example $f_p(z) = z^2+p$ (with $p \in \mathbb C$) \cite{Carleson}. It is associated with a fractal curve which is the border between the Fatou set of the values $z_0 \in \mathbb C$ having a bounded orbit $(z_n)_{n\in \mathbb N}$ (with $z_{n+1} = f_p(z_n)$) from the Julia set of the values with unbounded orbits. Reciprocally, another fractal, the Mandelbrot set, is the border between the values $p \in \mathbb C$ for which the orbit of $z_0=0$ is bounded from the values $p$ for which it is unbounded. The maps studied in \cite{Benchmann,Terno,Kiss1,Kiss2,Kiss3,Guan} and in this paper to represent the competition between decoherence and purification, belong to the family of the Julia map $f_p$.\\
Since the mixed state space is larger than the pure state space, the associated map describing the competition between decoherence and purification on a qubit has a phase space and a parameter space larger than $\mathbb C$. We will see that the map can be represented into the quaternion space $\mathbb H$. We can then think that the borders between the different behaviours are not simple fractal curves but more dimensional objects as Mandelbulbs (see \cite{Aron,Alonso}) or quaternionic Julia sets \cite{Norton}.\\
This paper is organised as follows. Firstly, we present the purification protocol. Second section presents the quaternionic representation of the qubit mixed states. Third section presents the quaternionic representation of the competition between decoherence and purification. Fifth section shows the results of this competition (with the fractal borders between the area dominated by the purification and the area dominated by the decoherence). Finally, we draw the quaternionic fractal sets resulting from the competition.

\section{The purification protocol}
Let $z \in \mathbb C$ be the complex parametrisation of a pure state of a qubit:
\begin{eqnarray}\label{purestate}
  |\psi \rangle & = & \frac{z|0\rangle + |1 \rangle}{\sqrt{1+|z|^2}} \\
  |\psi \rangle \langle \psi | & = & \frac{1}{1+|z|^2} \left(\begin{array}{cc} |z|^2 & z \\ \bar z & 1 \end{array} \right)
\end{eqnarray}
$z$ is the complex coordinates onto the Bloch sphere of the qubit states (the complex plane is the stereographic projection of the Bloch sphere). More precisely, $|\psi \rangle\langle \psi| = \left(\begin{array}{cc} p_0 & c \\ \bar c & p_1 \end{array} \right)$ where $p_0 = \frac{|z|^2}{1+|z|^2}$ is the probability of occupation of the state $|0\rangle$ when the qubit is in the state $|\psi\rangle$ (or in other words, if the qubit is in the state $|\psi \rangle$ and if we measure the value of the qubit, the probability to obtain the result $0$ is $p_0$). $p_1 = \frac{1}{1+|z|^2}$ is the probability of occupation of the state $|1\rangle$. $|c| = \frac{|z|}{1+|z|^2}$ is the coherence of the quantum state. If $|\psi \rangle$ is a true Schr\"odinger cat $|\psi \rangle = \frac{1}{\sqrt 2}(|0\rangle + e^{\iquat \phi} |1\rangle)$ then the coherence $|c| = \frac{1}{2}$ is maximal, indicating that the qubit is in a state furthest from the classical case. In contrast, if $|\psi \rangle = |0\rangle$, the coherence is zero, indicating that the qubit is in a state similar to a classical one (without quantum superposition the qubit has the behaviour of a classical bit). $\arg c = \arg z$ is the phase difference between $|0\rangle$ and $|1\rangle$, it is responsible of interference phenomena. For example, suppose that the qubit is initially in the state $|\psi\rangle = \frac{1}{\sqrt 2}(|0\rangle + |1\rangle)$ but after a transformation it becomes $|\psi'\rangle =  \frac{1}{\sqrt 2}(|0\rangle + e^{\iquat \phi} |1\rangle)$. These are two true Schr\"odinger cats (same probabilities of occupation and same coherence), but the survival probability of the initial state (the probability to recover the quantum behaviour of the initial state after the transformation) is $|\langle \psi|\psi' \rangle|^2 = \frac{1+\cos(\phi)}{2} <1$ (if $\phi \not=0$). This is due to the interferences between the states $|\psi\rangle$ and $|\psi'\rangle$.\\
The purification protocol $S$ studied in \cite{Benchmann,Terno,Kiss1,Kiss2,Kiss3,Guan} consists to the following algorithm (for the sake of simplicity, the states are not normalised in the presentation of the algorithm):
\begin{enumerate} \setcounter{enumi}{-1}
\item Initial state of the qubit: $z|0\rangle + |1\rangle$.
\item The state of the qubit is reproduced onto a second qubit (used only for the computation): $(z|0\rangle + |1\rangle) \otimes (z|0\rangle + |1\rangle) = z^2|00\rangle + z|01\rangle + z|10\rangle + |11\rangle$.
\item A \textit{controlled not} gate is applied onto the two qubits (entangling these ones): $z^2|00\rangle + z|01\rangle + z|10\rangle + |11\rangle \xrightarrow{CNOT} z^2|00\rangle + z|01\rangle + z|11\rangle + |10\rangle$.
\item A measure of the value of the second qubit is performed. The protocol succeeds if the measured value is $0$, in this case we select the first qubit. Otherwise, the protocol fails, the first qubit is rejected and it is necessary to restart: $z^2|00\rangle + z|01\rangle + z|11\rangle + |10\rangle \to z^2|00\rangle + |10\rangle = (z^2|0\rangle+|1\rangle) \otimes |0\rangle$.
\end{enumerate}
Formally, the protocol can be written as the following quantum operation:
\begin{equation}
  (\mathrm{id} \otimes |0\rangle \langle 0|) \mathbb U_{CNOT} (|\psi \rangle \otimes |\psi \rangle) \propto S|\psi\rangle \otimes |0\rangle
\end{equation}
($\propto$ stands for equal by definition up to a normalisation factor). $S$ induces the squaring of the pure state $|\psi\rangle \langle \psi|$:
\begin{equation}
  S|\psi\rangle = \frac{z^2|0\rangle + |1 \rangle}{\sqrt{1+|z|^4}}
\end{equation}
$S$ is not a logical gate (it is not a linear unitary operator), $S$ is a quantum information protocol which is nonlinear because of the  measurement onto the entangled second qubit and the post-selection of the first one depending on the result of the measurement. The nonlinearity results then from the gain of information by the measurement followed by the post-selection.\\

Let $U = e^{-\ihbar^{-1} H \Delta t} = \left(\begin{array}{cc} e^{\iquat \alpha} \cos x & e^{\iquat \varphi} \sin x \\ -e^{-\iquat \varphi} \sin x & e^{-\iquat \alpha} \cos x \end{array} \right)$ be the evolution operator of the qubit during a short time duration $\Delta t$. $H=\frac{\hbar}{2} (\omega \sigma_z + \RE(b) \sigma_x + \IM(b) \sigma_y)$ is the qubit quantum Hamiltonian, with $\tan x = \frac{|b|\sin(r\Delta t/2)}{\sqrt{|b|^2\cos^2(r\Delta t/2)+\omega^2}}$, $\tan \alpha = -\frac{\omega}{r} \tan(r \Delta t/2)$, $\varphi = \arg b - \frac{\pi}{2}$ and $r=\sqrt{\omega^2+|b|^2}$. $\pm \frac{\hbar \omega}{2}$ are the energies of the two states $|0\rangle$ and $|1\rangle$ (with the gauge choice concerning the energy origin such that $\tr H=0$). $b$ is a constant external field coupling the two qubit states. For example, if the qubit is physically realised by a $1/2$-spin (a quantum magnetic moment), $\hbar \omega$ is the energy split by Zeeman effect induced by a constant magnetic field in the $z$-direction, and $\RE(b) \vec e_x + \IM(b)\vec e_z$ is a transverse constant magnetic field acting on the spin. $\pm \frac{\hbar}{2} r$ are the eigenenergies in presence of the external magnetic field. $U$ defines the evolution of the qubit under the external field applied during $\Delta t$. In the context of a quantum computer, $U$ can be viewed as a single qubit logical gate, $\Delta t$ being the time needed to apply this one.\\

The succession of the purification protocol and of the evolution operator induces on a pure qubit state the following transformation:
\begin{equation}
  US|\psi\rangle = \frac{f_{\alpha,p}(z)|0\rangle + |1 \rangle}{\sqrt{1+|f_{\alpha,p}(z)|^2}}
\end{equation}
with the complex map:
\begin{equation} \label{mapComplex}
  f_{\alpha,p}(z) = \frac{z^2 e^{\iquat \alpha} + p}{e^{-\iquat \alpha} - \bar p z^2}
\end{equation}
$p = e^{\iquat \varphi} \tan x$. $f_{\alpha,p}$ is similar to a ``renormalised'' Julia map. The iteration of protocols $US$ with interval $\Delta t$, i.e.
\begin{equation}
  (US)^n|\psi \rangle = USUS...US|\psi\rangle \equiv \frac{z_n|0\rangle + |1\rangle}{\sqrt{1+|z_n|^2}}
\end{equation}
is then represented by the dynamical system $z_{n+1} = f_{\alpha,p}(z_n)$. It as been studied in \cite{Kiss1,Kiss2,Kiss3} (with $\alpha \in 2\pi \mathbb Z$) and in \cite{Guan} (with $\alpha \not\in 2\pi \mathbb Z$).

\section{Quaternionic representation}
A pure state $|\psi \rangle \langle \psi| = \left(\begin{array}{cc} p_0 & c \\ \bar c & p_1 \end{array} \right)$ eq.(\ref{purestate}) is associated with an isolated qubit. This state satisfies $\tr(|\psi\rangle \langle \psi|^2)=1$ (purity equal to 1) or equivalently $|c|^2=p_0p_1$. But in the reality, the qubit is submitted to environment noises. To simply the discussion here, we suppose that the effects of these noises can be modelled by a random process onto the qubit state (in a pure quantum model, where the environment is modelled by a large quantum system, the effect of the noises are in fact an entanglement between the qubit and the environment, but the results are the same than with a random process, see \cite{Breuer}). We can then write that
\begin{equation}
  |\psi(\{\omega\})\rangle = a_0(\{\omega\}) |0\rangle + a_1(\{\omega\})|1\rangle
\end{equation}
where $a_0$ and $a_1$ are complex numbers (such that $|a_0|^2+|a_1|^2=1$) depending on random variables $\{\omega\}$ associated with the environment noises. The qubit state accessible to the experimentalist (who cannot control the random process) is then the density matrix
\begin{equation}
  \rho = \mathbb E\left(|\psi(\{\omega\})\rangle \langle \psi(\{\omega\})|\right)
\end{equation}
where $\mathbb E$ stands for the average with respect to the random process associated with the environment noises (if the environment is modelled by a large quantum system, $\mathbb E$ is replaced by the partial trace over the environment quantum degrees of freedom). Anew $\rho = \left(\begin{array}{cc} p_0 & c \\ \bar c & p_1 \end{array} \right)$ where $p_0 = \mathbb E(|a_0|^2)$ is the probability of occupation of the state $|0\rangle$ (average probability to find the qubit with the value $0$ if we measure this one, the average being onto the random process). $p_1 = \mathbb E(|a_1|^2)$ is the probability of occupation of the state $|1\rangle$. $|c| = |\mathbb E(a_0 \bar a_1)|$ is the coherence, but now $\tr(\rho^2)<1$ (the purity is smaller than 1, since $\mathbb E(|\psi\rangle \langle \psi|^2) \not= \left(\mathbb E(|\psi\rangle \langle \psi|)\right)^2$) or equivalently $|c|< \sqrt{p_0p_1}$ (the effect of the environment noises are called decoherence since the coherence falls). For example, consider the pure state $\rho = \left(\begin{array}{cc} 1/2 & 1/2 \\ 1/2 & 1/2 \end{array}\right)$ without noise, and $\rho' = \left(\begin{array}{cc} 1/2 & 0 \\ 0 & 1/2 \end{array}\right)$ a density matrix for which the coherence has fallen to $0$ under the effect of the noises. The probabilities of occupation are $1/2$ in the two cases. For the first case, the coherence is maximal and then the state is furthest from the classical case (it is a single true Schr\"odinger cat state $\frac{1}{\sqrt 2}(|0\rangle + |1\rangle)$). For the second case, the coherence is zero, meaning that the density matrix corresponds to classical state for which the probability to the state be $|0\rangle$ and the one to the state be $|1\rangle$ are $1/2$ (the state is unknown due to the random process). The mixed state $\rho'$ is then a statistical mixture of two classical states (the state can be $|0\rangle$ \textbf{or} $|1\rangle$) whereas the pure state $\rho$ is a quantum superposition of two states (the state is both $|0\rangle$ \textbf{and} $|1\rangle$). We can note the difference between the purity and the coherence. A pure state ($\tr(\rho^2)=1$) means a state without statistical uncertainty (a state without unknown information due to the noises). A state with maximal coherence means a state with maximal quantum superposition, so a state with strong quantum behaviour. A state can be pure with coherence zero, as for example $|0\rangle \langle 0|$. In general, a mixed state with $c\not=0$ represents a state with both quantum superposition and statistical mixture.\\
As for the pure states, we want to parametrise the mixed states with a complex variable $z$. But we need to add a second parameter to describe the coherence since $|c|\not=\sqrt{p_1p_2}$ that we call the mixing angle $\lambda$: $\cos \lambda \equiv \frac{|c|}{\sqrt{p_1p_2}}$. The mixed state of the qubit after the parametrisation is then the following density matrix:
\begin{equation}
  \rho = \frac{1}{1+|z|^2} \left(\begin{array}{cc} |z|^2 & z \cos \lambda \\ \bar z \cos \lambda & 1 \end{array} \right)
\end{equation}
$\lambda$ is the mixing angle, for $\lambda=0$ $\rho$ is pure state and for $\lambda = \frac{\pi}{2}$ the coherence of the qubit is zero (maximal mixing). It needs to take into account this new parameter in the representation.\\
The purification protocol can be performed onto a mixed state (with the same algorithm) as for a pure state:
\begin{equation}
  (\mathrm{id} \otimes |0\rangle \langle 0|) \mathbb U_{CNOT} (\rho \otimes \rho) \mathbb U_{CNOT} (\mathrm{id} \otimes |0\rangle \langle 0|) \propto S(\rho) \otimes |0\rangle \langle 0|
\end{equation}
We have then $\lim_{n\to +\infty} S^n(\rho) = |0\rangle \langle 0|$ if $|z|>1$ or $|1\rangle \langle 1|$ if $|z|<1$ (where $S^n$ stands for the application of $S$ $n$ times). $S^n$ (with $n$ large) transforms then a mixed state $\rho$ to a pure state $|0\rangle \langle 0|$ or $|1\rangle \langle 1|$. This is the reason for which $S$ is called purification protocol. When $S$ acts alone, it purifies mixed states to pure states without coherence (without quantum superposition). The role of the logical gate (or of the evolution induced by an external field) $U$ is to permit to reach fixed points or cycles of pure states with a non-zero coherence \cite{Portiko}. But the dynamics induced by $(US)^n$ becomes chaotic as shown in \cite{Benchmann,Terno,Kiss1,Kiss2,Kiss3,Guan}. Moreover $(US)^n$ is an idealisation for which the environment noises are turned off during the application of the protocol. In the realistic situations, at each iteration noises induce decoherence onto the qubit. The goal of this paper is to study the competition between the purification effect of the protocol $S$ and the decoherence effect of the noises.\\

Due to the added parameter $\lambda$ needed to define a mixed state, we cannot represent this one by a single complex number. But we can think that this is possible with a single quaternionic number. In \cite{Mosseri} the authors introduce a quaternionic representation of qubit pair states in order to study the entanglement phenomenon. The quaternion space $\mathbb H$ is the set of noncommutative numbers $\zeta = a + \iquat b + \jquat c + \kquat d$, with $a,b,c,d \in \mathbb R$, $\iquat^2 = \jquat^2 = \kquat^2 = -1$ and $\iquat \jquat = \kquat$, $\jquat \iquat = -\kquat$, $\jquat \kquat = \iquat$, $\kquat \jquat = -\iquat$, $\kquat\iquat = \jquat$, $\iquat \kquat = -\jquat$. We denote: $\RE(\zeta)=a$, $\IM_1(\zeta)=b$, $\IM_2(\zeta)=c$, $\IM_3(\zeta)=d$, $\CO(\zeta) = a + \iquat b$ and $|\zeta|^2 = \zeta \bar \zeta = a^2+b^2+c^2+d^2$. Note that $\zeta^{-1} = \frac{\bar \zeta}{|\zeta|^2}$. For a state of two qubits:
\begin{equation}
  |\Psi \rangle = \frac{z \cos\lambda_0 |00\rangle + z \sin \lambda_0 |01 \rangle + \cos \lambda_1 |10\rangle + \sin \lambda_1 |11 \rangle}{\sqrt{1 + |z|^2}}
\end{equation}
with $z \in \mathbb C$, $\lambda_i \in [0,2\pi]$, the quaternionic representation is
\begin{equation}
  (\zeta_0,\zeta_1) = (z e^{\jquat \lambda_0}, e^{\jquat \lambda_1}) \in \mathbb H^2
\end{equation}
The mixed state of the first qubit (the mixing resulting from the entanglement with the second one) is then
\begin{eqnarray}
  \rho & = & \tr_2 |\Psi \rangle \langle \Psi| \\
  & = & \frac{1}{1+|z|^2} \left(\begin{array}{cc} |z|^2 & z \cos(\lambda_0-\lambda_1) \\ \bar z \cos(\lambda_0-\lambda_1) & 1 \end{array} \right)\\
  & = & \frac{1}{1+|z|^2} \CO \left(\begin{array}{cc} |z|^2 & z e^{\jquat (\lambda_0-\lambda_1)} \\ \bar z e^{\jquat (\lambda_0-\lambda_1)} & 1 \end{array} \right)
\end{eqnarray}
where $\tr_2$ is the partial trace onto the state space of the second qubit. The density matrix can be then represented by the quaternionic number $\zeta = z e^{\jquat \lambda} \in \mathbb H$ (with $\lambda=\lambda_0-\lambda_1$ for the entanglement case) with
\begin{equation} \label{quaterrep}
  \rho = \frac{1}{1+|\zeta|^2} \CO \left(\begin{array}{cc} |\zeta|^2 & \zeta \\ \bar \zeta & 1 \end{array} \right)
\end{equation}
Note that $\zeta = z e^{\jquat \lambda} = z\cos \lambda + \jquat \bar z \sin \lambda = e^{\iquat \phi} (C - \frac{1}{2} \mathcal C \jquat)$ where $\phi = \arg z$ is the phase, $C = |z|\cos(\lambda)$ is the coherence of the first qubit, and $\mathcal C = -2|z|\sin(\lambda_0-\lambda_1)$ is the concurrence of the entanglement between the two qubits \cite{Bengtsson}.\\
We adopt the quaternionic representation of the density matrix eq. \ref{quaterrep} also for mixed states resulting from a decoherence process (note that any qubit mixed state can be represented by an entangled state of the qubit with an ancilla qubit, by using the Schmidt purification procedure \cite{Bengtsson}). For $\zeta \in \mathbb H$, $\frac{|\zeta|^2}{1+|\zeta|^2}$ is the population of the state $|0\rangle$ and $|\CO(\zeta)|$ is the coherence of the mixed state. With these interpretations, several $\zeta$ in $\mathbb H$ correspond to the same mixed state, it can be then interesting to transform any $\zeta$ in the form $ze^{\jquat \lambda}$:
\begin{equation}
  \mathrm{p}(\zeta) = \begin{cases} \CO(\zeta)+ \frac{|\zeta-\CO(\zeta)|}{|\CO(\zeta)|}\CO(\zeta) \jquat = ze^{\jquat \lambda} & \text{if } \CO(\zeta) \not=0 \\
    \zeta = z e^{\jquat \frac{\pi}{2}} & \text{if } \CO(\zeta)=0
  \end{cases}
\end{equation}
with $z = \CO(\zeta)$ and $\cos \lambda = \frac{|\CO(\zeta)|}{|\zeta|}$ for the case $\CO(\zeta)\not=0$. 

\section{Dynamics in the quaternionic representation}
\begin{figure}
  \includegraphics[width=7cm]{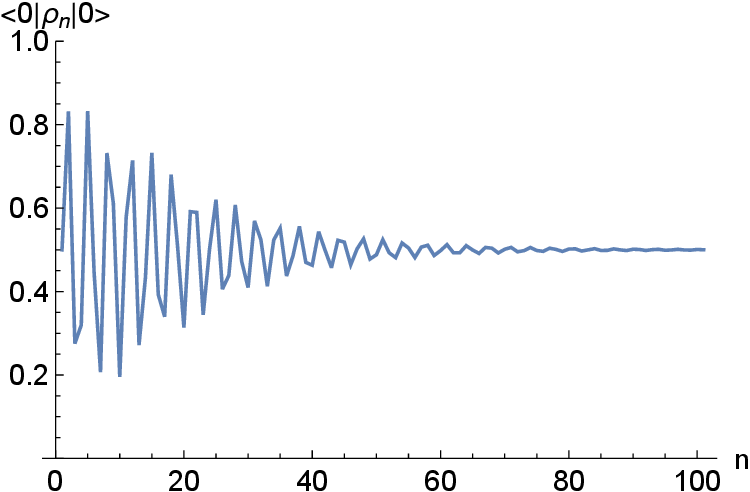}\\
  \includegraphics[width=7cm]{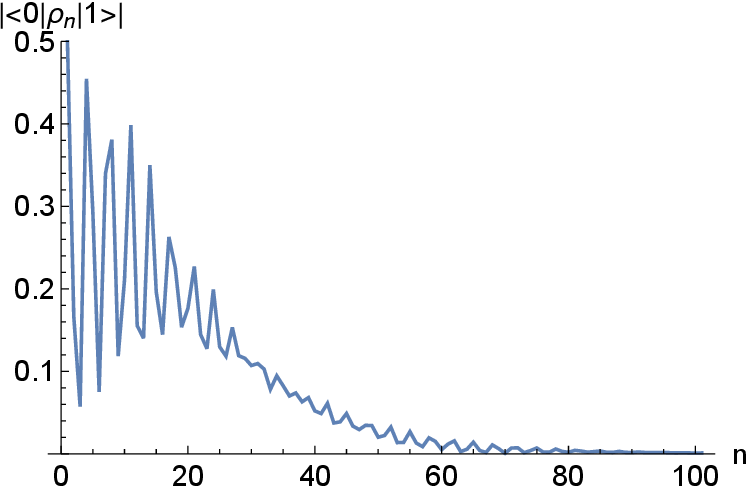}\\
  \includegraphics[width=7cm]{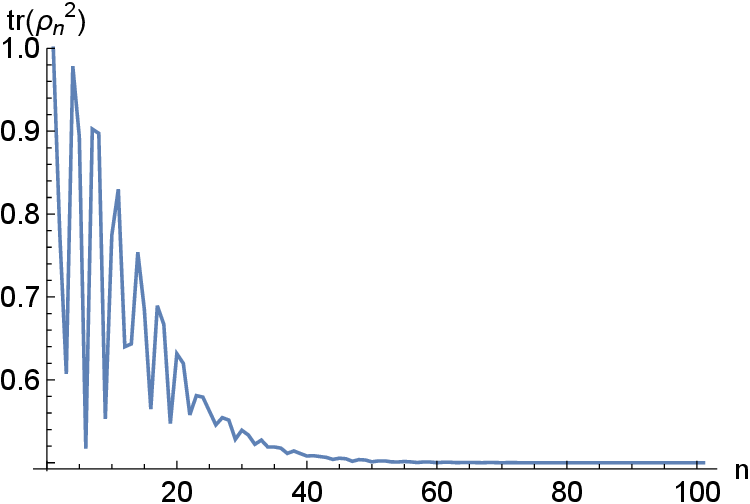}\\
  \caption{\label{dugraph} The dynamical system $\zeta_{n+1} = \mathrm{du}(\zeta_n)$ with $\mathrm{du}$ defined by eq. \ref{du}, with $\zeta_0 = 1$, $\alpha=0.1$, $\beta=\gamma=0$ and $q=1+\kquat$. Up: population $\langle 0|\rho_{n}|0\rangle = \frac{|\zeta_n|^2}{1+|\zeta_n|^2}$; middle: coherence $|\langle 0|\rho_n|1\rangle|=|\CO \zeta_n|$, down: purity $\tr (\rho_n^2) = \frac{|\zeta_n|^4+2|\CO\zeta_n|^2+1}{(1+|\zeta_n|^2)^2}$.}
\end{figure}

We want to consider transformations $DUS(\rho)$ where $S$ is the purification protocol, $U$ is the evolution operator map of the qubit, and $D$ is a decoherence process ($DU$ can come from the integration of a Lindblad equation during $\Delta t$, see \cite{Breuer}). The purification protocol induces the squaring of the density matrix:
\begin{equation}
  S(\rho) = \frac{1}{1+|z|^4} \left(\begin{array}{cc} |z|^4 & z^2 \cos^2 \lambda \\ \bar z^2 \cos^2 \lambda & 1 \end{array} \right)
\end{equation}
Let $\mathrm{s}: \mathbb H \to \mathbb H$ be the map such that
\begin{equation}
  \frac{1}{1+|\mathrm{s}(\zeta)|^2} \CO \left(\begin{array}{cc} |\mathrm{s}(\zeta)|^2 & \mathrm{s}(\zeta) \\ \mathrm{s}(\bar \zeta) & 1 \end{array} \right) = S(\rho)
\end{equation}
Unfortunately, $\mathrm{s}$ is more complicated than a square power:
\begin{equation}
  \mathrm{s}(\zeta) = (\CO \zeta)^2 + \jquat \IM_2 \left((\zeta-\CO\zeta) \CO\zeta \right) + \kquat \frac{|\zeta|^2 \IM_2 \zeta}{|\RE \zeta+\jquat \IM_2 \zeta|}
\end{equation}
The evolution of the density matrix (evolution between two purifications due to an external field or to a logical gate without decoherence processes) is defined with the evolution operator $U$ as in section 2 by
\begin{equation}
  U(\rho) = U \rho U^\dagger
\end{equation}
it corresponds to the map $\mathrm{u}:\mathbb H \to \mathbb H$
\begin{equation} \label{evolH}
  \mathrm{u}(\zeta) = (e^{\iquat \alpha} \zeta + p)(e^{-\iquat \alpha} - \bar p \zeta)^{-1}
\end{equation}
with $p = e^{\iquat \varphi} \tan x \in \mathbb C$.\\
For the decoherence processes, we can consider pure dephasing processes \cite{Marquardt}:
\begin{equation}
  D(\rho) = \frac{1}{1+|z|^2} \left(\begin{array}{cc} |z|^2 & (1-\beta)z \cos \lambda \\ (1-\beta)\bar z \cos \lambda & 1 \end{array} \right)
\end{equation}
with $0<\beta<1$ the decoherence rate during $\Delta t$. If $\beta \ll 1$, $(1-\beta) \cos \lambda = \cos \lambda'$ with $\lambda' = \lambda + \beta \mathrm{cotan}\, \lambda + \mathcal O(\beta^2)$. It follows that the decoherence corresponds to the map $\mathrm{d}:\mathbb H \to \mathbb H$
\begin{equation}
  \mathrm{d}(\zeta) = \begin{cases} \zeta e^{\jquat \frac{|\CO\zeta|}{|\zeta-\CO\zeta|} \beta} & \text{if } |\zeta-\CO\zeta|\not=0 \\ \zeta e^{\jquat \sqrt{2\beta}} & \text{else} \end{cases}
\end{equation}
which is a dephasing (of the second kind) in $\mathbb H$. The map $f_{\alpha,\beta,p}(\zeta) = \mathrm{dus}(\zeta)$ is a generalisation in $\mathbb H$ of the Julia map, it induces a dynamical system in $\mathbb H$, $\zeta_{n+1} = f_{\alpha,\beta,p}(\zeta_n)$, corresponding to a competition between the pure dephasing process and the iterated purification protocol.\\
Another example of decoherence process consists to consider the natural generalisation of the map (\ref{evolH}), $\mathrm{du}: \mathbb H \to \mathbb H$, with
\begin{equation} \label{du}
  \mathrm{du}(\zeta) = (e^{\iquat \alpha} e^{\jquat \beta} e^{\kquat \gamma}  \zeta + q)(e^{-\kquat \gamma} e^{-\jquat \gamma} e^{-\iquat \alpha} - \bar q \zeta)^{-1}
\end{equation}
with $q \in \mathbb H$. This map induces a dynamics with decoherence as we can see it figure \ref{dugraph}.
Finally, the map $f_{\alpha,\beta,\gamma,q}(z) = \mathrm{pdus}(z)$ defines a generalisation in $\mathbb H$ of the map (\ref{mapComplex}) representing a competition between a decoherence process and the purification protocol.

\section{Results of the competition}
\begin{figure}
  \includegraphics[width=6cm]{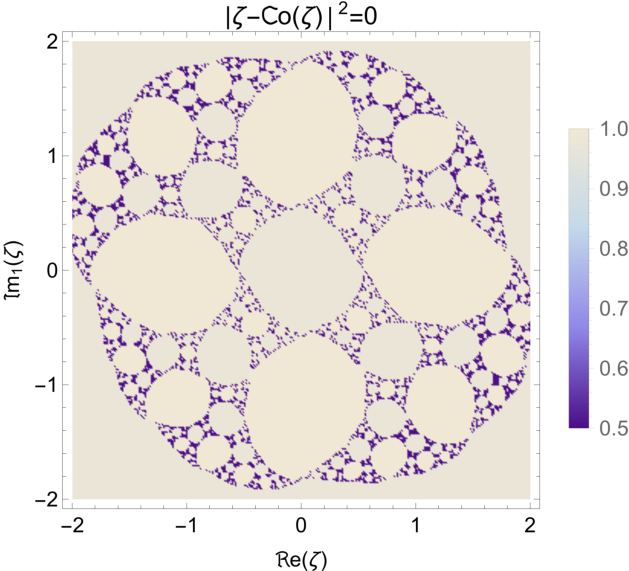}\includegraphics[width=6cm]{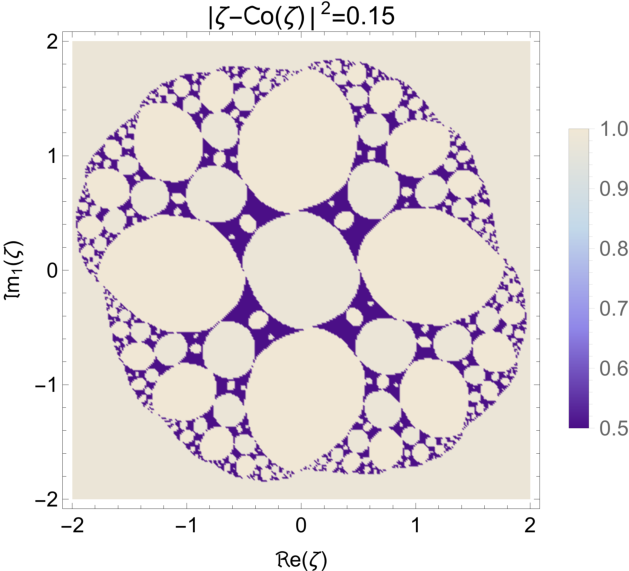}\\
  \includegraphics[width=6cm]{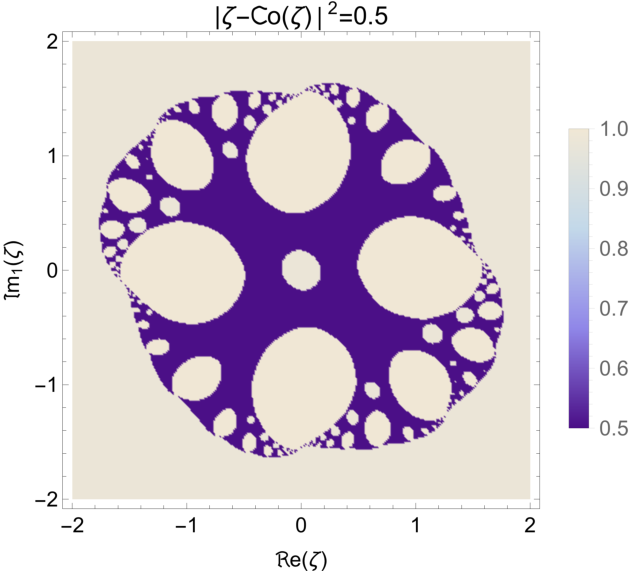}\includegraphics[width=6cm]{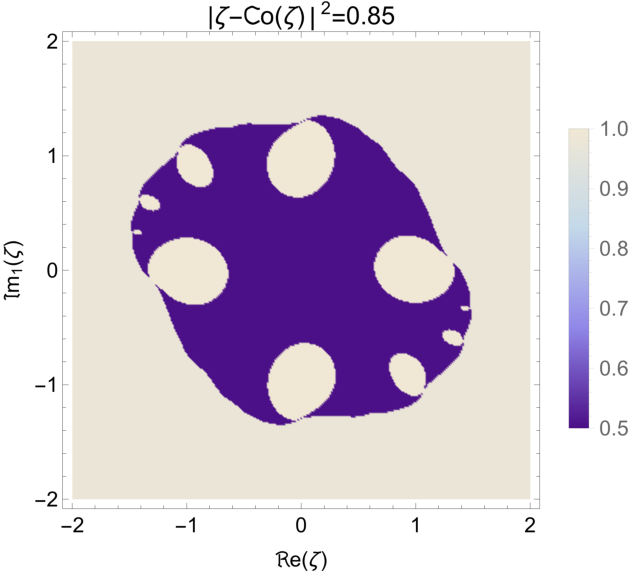}\\
  \caption{\label{purityfractaldephasing} Purity $\tr(\rho_N^2) = \frac{|\zeta_N|^4+2|\CO\zeta_N|^2+1}{(1+|\zeta_N|^2)^2}$ ($N=100$) for the dynamical system $\zeta_{n+1} = f_{\alpha,\beta,p}(\zeta_n)$ ($\alpha=0$, $\beta=0.01$, $p=1+0.1\iquat$) corresponding to a competition between the purification protocol and a pure dephasing decoherence process. The planes represent the initial condition $\zeta_0 = z_0e^{\jquat \lambda_0}$ (with $|\zeta_0-\CO\zeta_0|^2=C^{st}$) coloured with respect to the purity at ``the end'' of its orbit.}
\end{figure}

\begin{figure}
  \includegraphics[width=6cm]{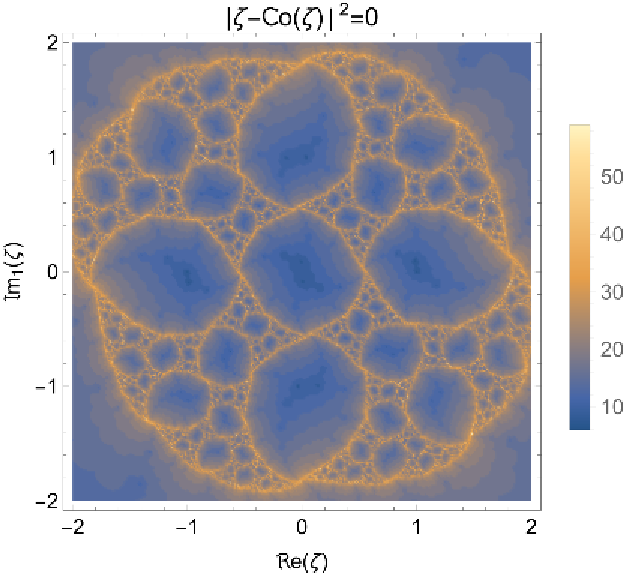}\includegraphics[width=6cm]{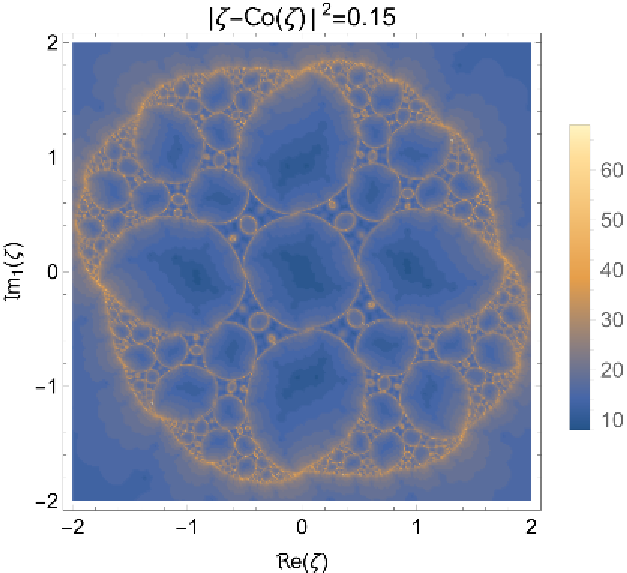}\\
  \includegraphics[width=6cm]{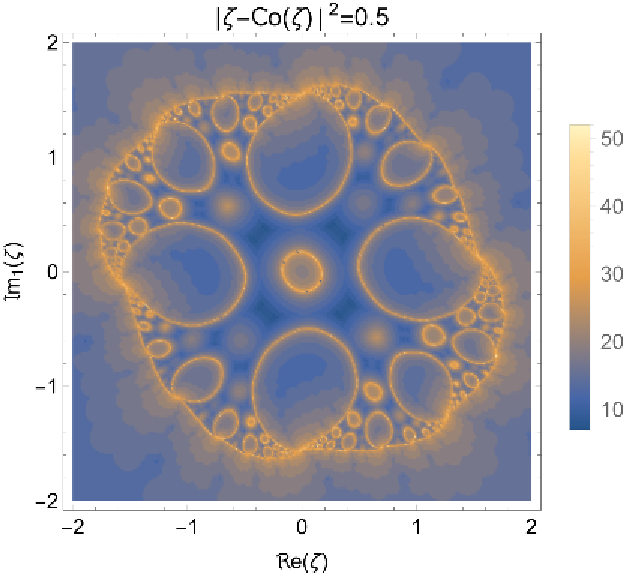}\includegraphics[width=6cm]{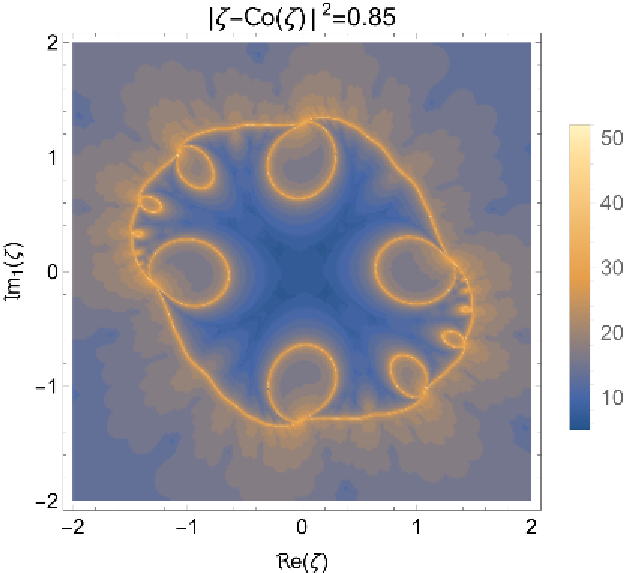}\\
  \caption{\label{stabilityfractaldephasing} For the dynamical system $\zeta_{n+1} = f_{\alpha,\beta,p}(\zeta_n)$ ($\alpha=0$, $\beta=0.01$, $p=1+0.1\iquat$) corresponding to a competition between the purification protocol and a pure dephasing decoherence process, the number of iterations needed to reach a cycle (of period lower than 5). The planes represent the initial condition $\zeta_0 = z_0e^{\jquat \lambda_0}$ (with $|\zeta_0-\CO\zeta_0|^2=C^{st}$). The precision for the criterion of return after one period is chosen to be $10^{-4}$.}
\end{figure}

\begin{figure}
  \includegraphics[width=6cm]{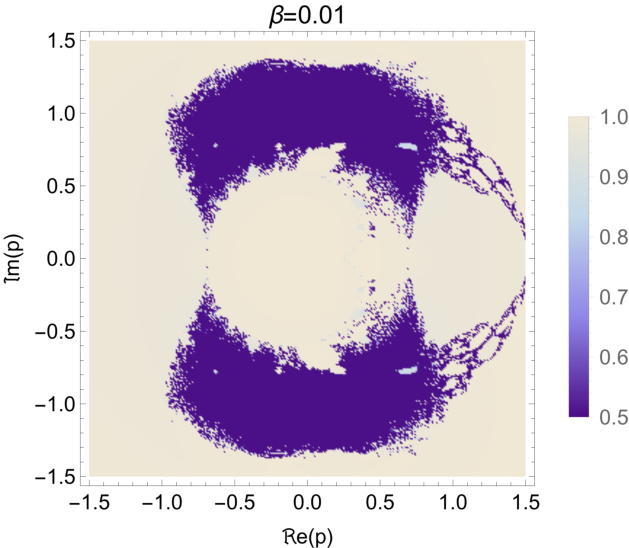}\includegraphics[width=6cm]{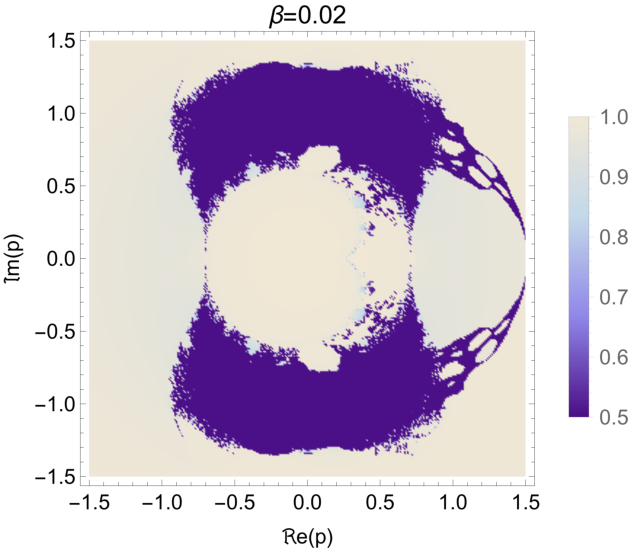}\\
  \includegraphics[width=6cm]{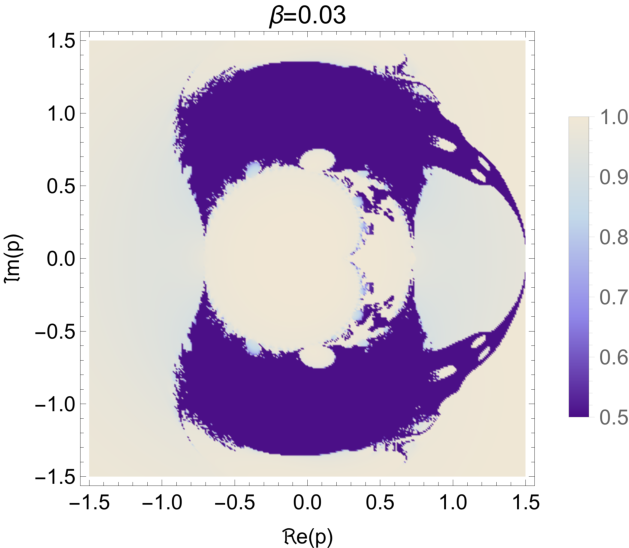}\includegraphics[width=6cm]{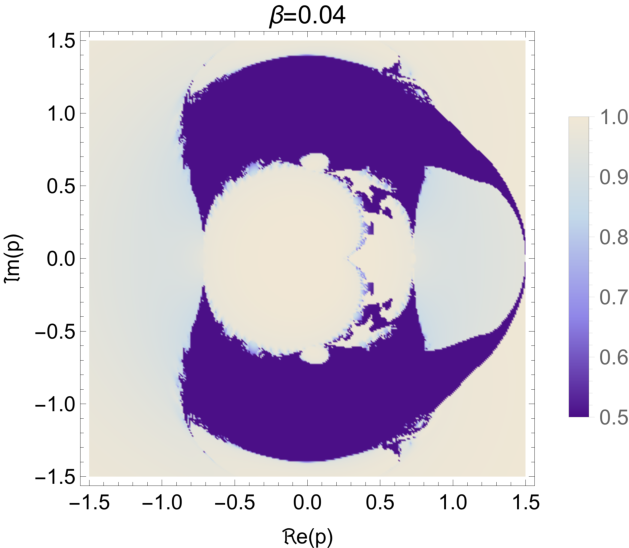}\\
  \caption{\label{purityMandeldephasing} Purity $\tr(\rho_N^2) = \frac{|\zeta_N|^4+2|\CO\zeta_N|^2+1}{(1+|\zeta_N|^2)^2}$ ($N=100$) for the dynamical system $\zeta_{n+1} = f_{\alpha,\beta,p}(\zeta_n)$ ($\alpha=0$, $\zeta_0=0$) corresponding to a competition between the purification protocol and a pure dephasing decoherence process. The planes represent the parameter $p\in \mathbb C$ coloured with respect to the purity at ``the end'' of the corresponding orbit. Different values of $\beta$ are considered.}
\end{figure}

\begin{figure}
  \includegraphics[width=6cm]{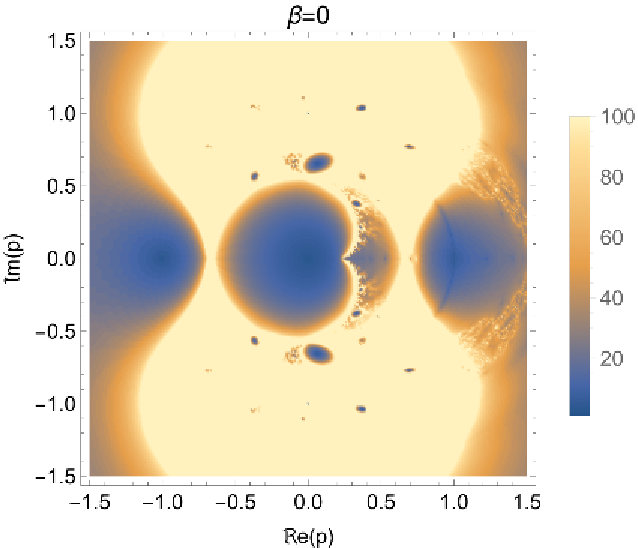}\includegraphics[width=6cm]{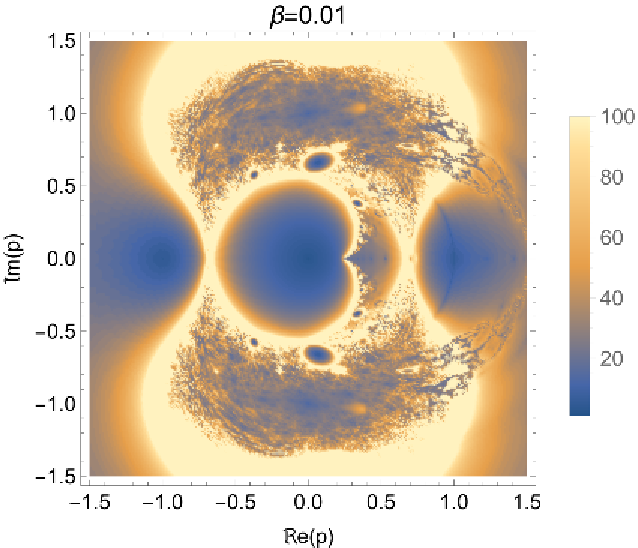}\\
  \includegraphics[width=6cm]{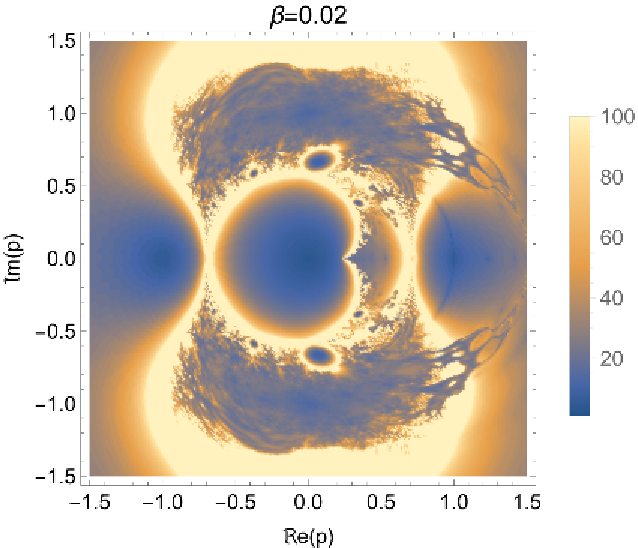}\includegraphics[width=6cm]{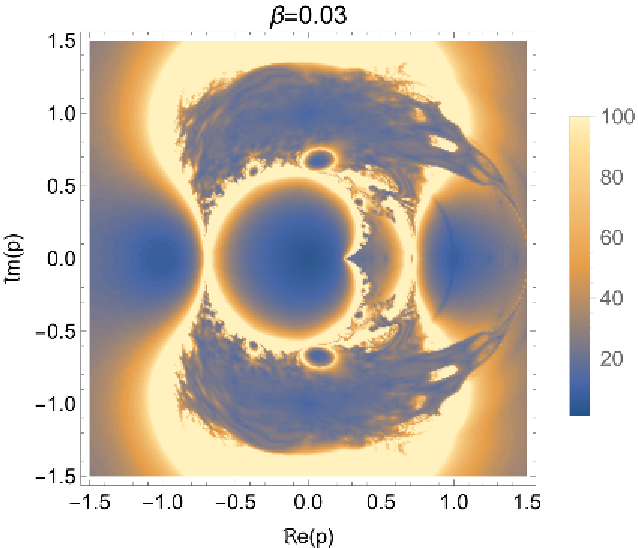}\\
  \caption{\label{stabilityMandeldephasing} For the dynamical system $\zeta_{n+1} = f_{\alpha,\beta,p}(\zeta_n)$ ($\alpha=0$, $\zeta_0=0$) corresponding to a competition between the purification protocol and a pure dephasing decoherence process, the number of iterations needed to reach a cycle (of period lower than 5). The planes represent the parameter $p\in \mathbb C$, different values of $\beta$ are considered. The precision for the criterion of return after one period is chosen to be $10^{-4}$.}
\end{figure}

The instability of the purification protocol which induces fractal borders between bounded and unbounded orbits in the pure state space, involves also complicated behaviours in the competition between the purification protocol and the decoherence process. The border between states for which the purification wins ($\lim_{n \to + \infty} \tr (\rho_n^2) \simeq 1$) and for which the decoherence wins ($\lim_{n \to +\infty} \tr(\rho_n^2) \simeq 0.5$) is irregular with a highly fractal character in the neighbourhood of the pure states, see fig. \ref{purityfractaldephasing}.
The states in the border between the area dominated by the purification and the area dominated by the decoherence are instable in sense that in contrast with the states inside the two areas, their orbits do not reach cyclic points. We can see this fig. \ref{stabilityfractaldephasing}.
In the area dominated by the decoherence, the orbits reach fixed points (1-period cycles) with $\lambda=\frac{\pi}{2}$. In the area dominated by the purification, we find cyclic points as for the map without decoherence. These fractal curves are equivalent to the Julia set, but we can also consider the equivalent of the Mandelbrot set, i.e. the purity for a long time of the orbit of $z_0=0$ (fig. \ref{purityMandeldephasing}) and the stability of the orbit of $z_0=0$ (fig. \ref{stabilityMandeldephasing}).

\begin{figure}
  \includegraphics[width=6cm]{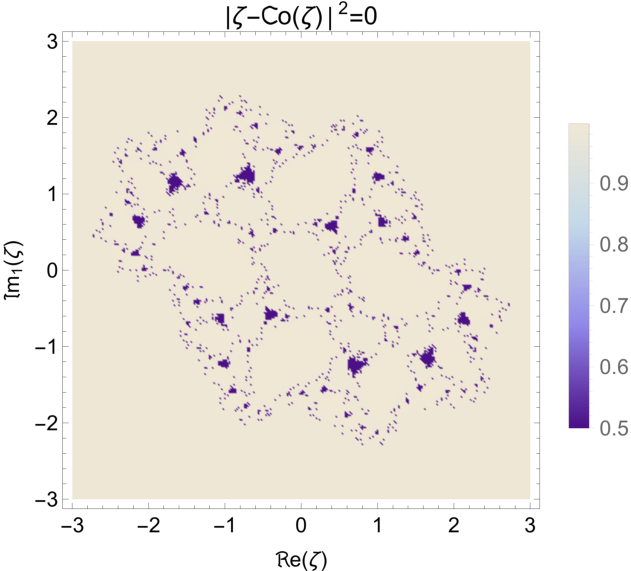}\includegraphics[width=6cm]{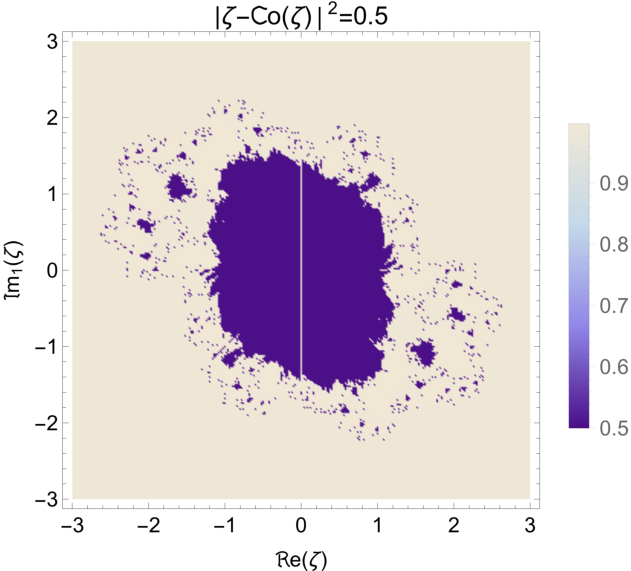}\\
  \includegraphics[width=6cm]{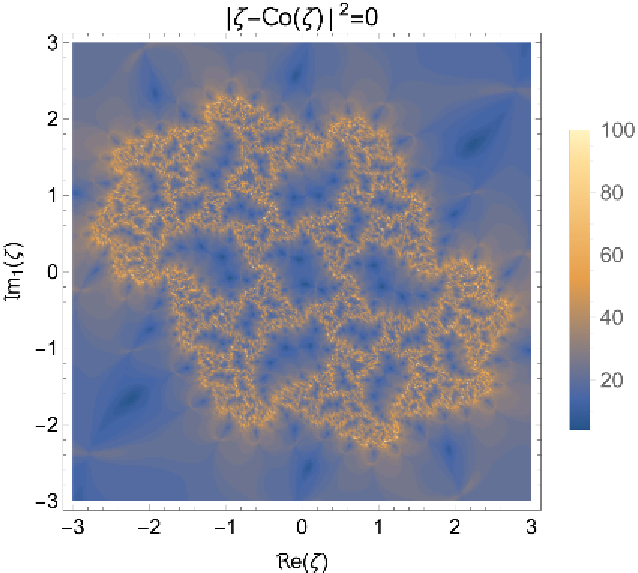}\includegraphics[width=6cm]{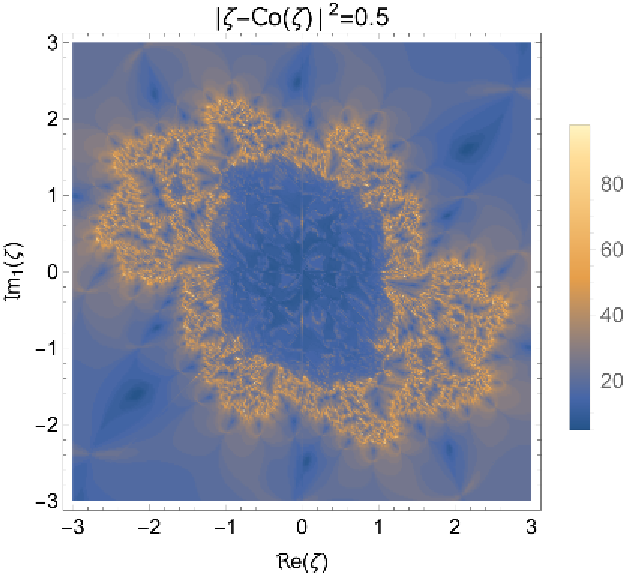}\\
  \caption{\label{purityfractalquat} Same as fig. \ref{purityfractaldephasing} (up) and \ref{stabilityfractaldephasing} (down) for the dynamical system $\zeta_{n+1} = f_{\alpha,\beta,\gamma,q}(\zeta_n)$ with $\alpha=0.1$, $\beta=0$, $\gamma=0$, $q=1+0.1\kquat$.}
\end{figure}

\begin{figure}
 \includegraphics[width=6cm]{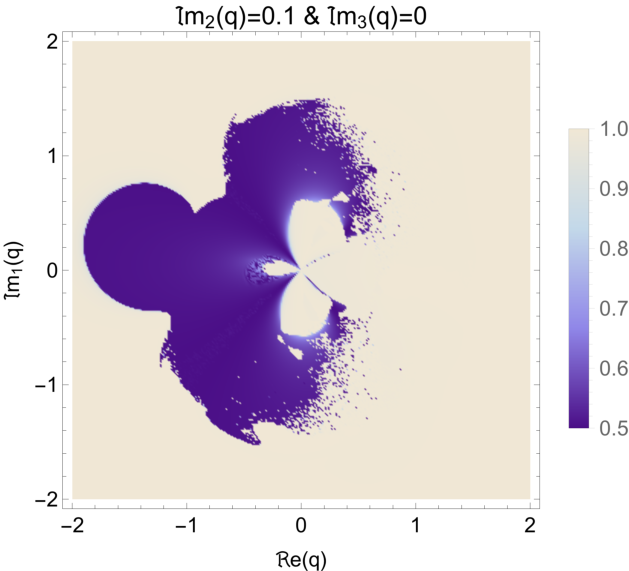}\includegraphics[width=6cm]{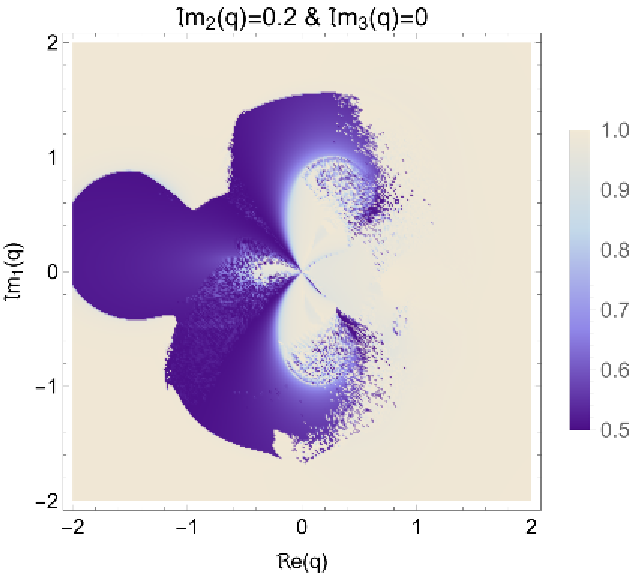}\\
 \includegraphics[width=6cm]{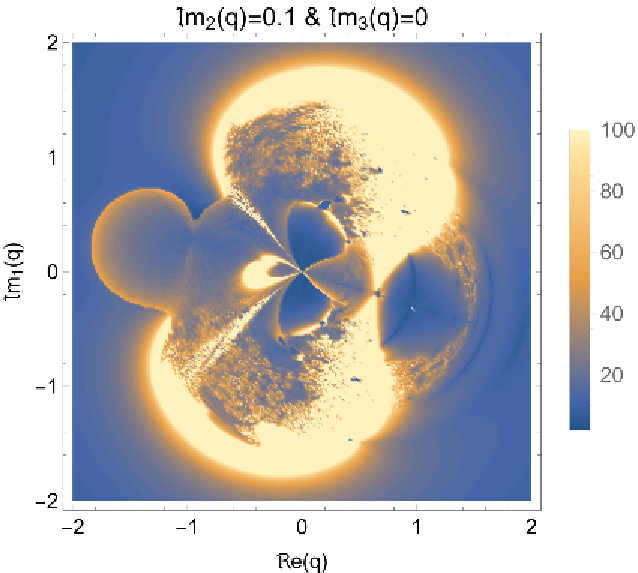}\includegraphics[width=6cm]{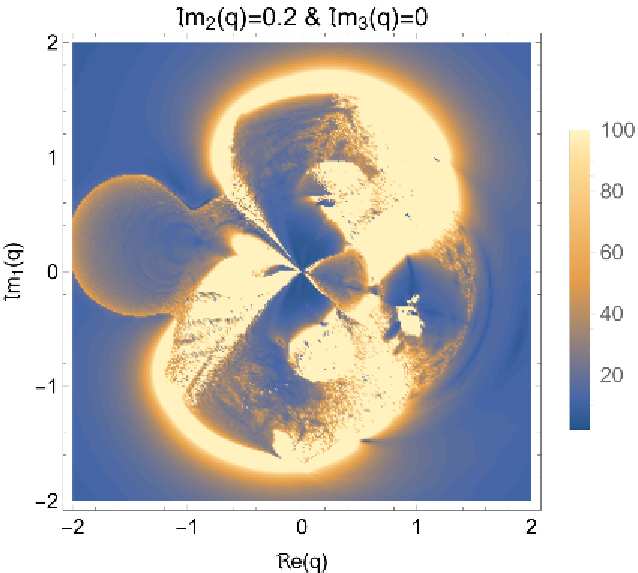}\\
  \caption{\label{stabilityMandelquat} Same as fig. \ref{purityMandeldephasing} (up) and \ref{stabilityMandeldephasing} (down) for the dynamical system $\zeta_{n+1} = f_{\alpha,\beta,\gamma,q}(\zeta_n)$ with $\alpha=0.1$, $\beta=0$, $\gamma=0$, and $\zeta_0=0$.}
\end{figure}

The observed behaviour is not dependent of the chosen particular decoherence process (pure dephasing). We recover it, but with another fractals, with the decoherence process defined by eq. \ref{du}, as we see it fig. \ref{purityfractalquat} \& \ref{stabilityMandelquat}.
For this decoherence process, the fixed point reached in the area dominated by the decoherence is the microcanonical distribution $\rho=\frac{1}{2}\mathrm{id}$ ($\zeta=\jquat$).

\section{Quaternionic fractal sets}
\begin{figure}
  \includegraphics[width=6cm]{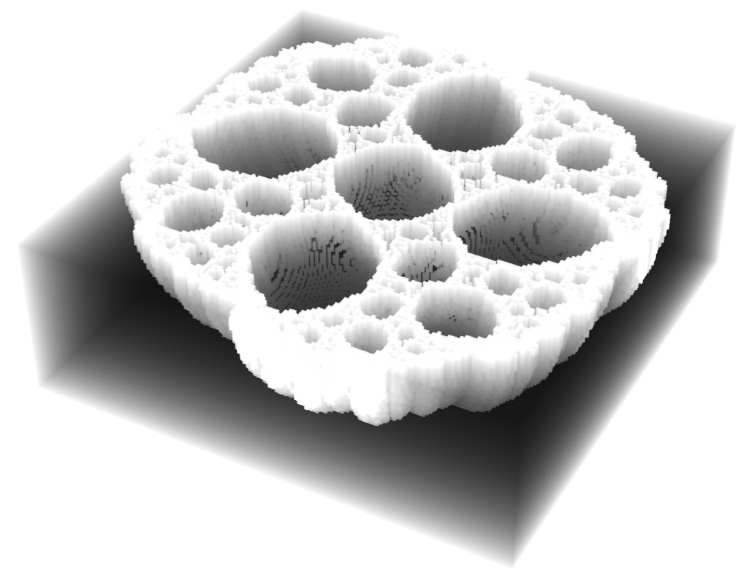} \includegraphics[width=6cm]{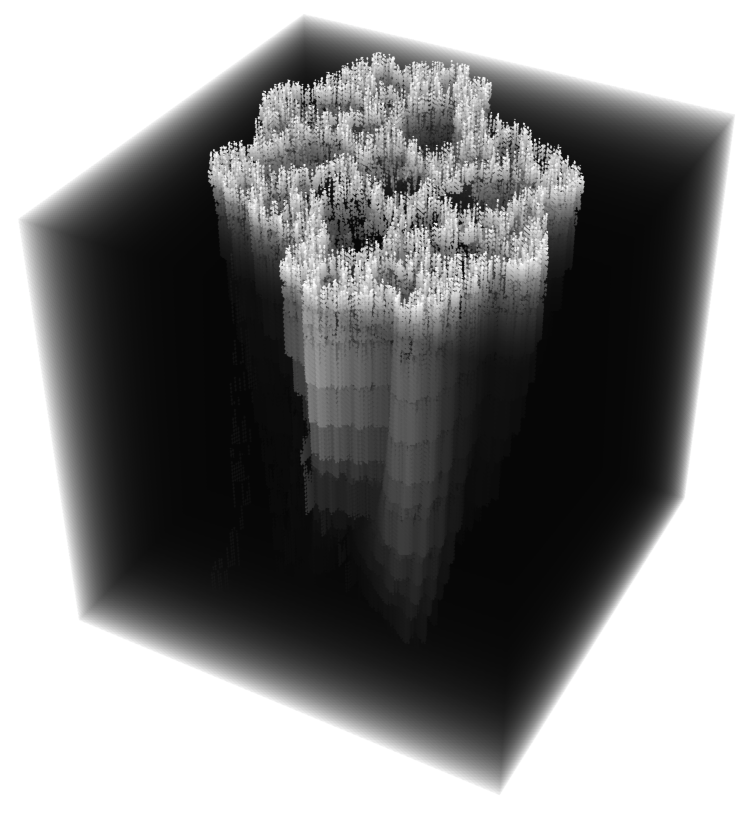}
  \caption{\label{Mandelbulb} Quaternionic fractal borders between the area dominated by the purification and the area dominated by the decoherence in the space $\mathbb R^3$ spanned by $(\RE(\zeta),\IM_1(\zeta),-|\zeta-\CO(\zeta)|)$ for the pure dephasing process (left) and the decoherence process eq. \ref{du} (right).}
\end{figure}

In the previous section, we have drawn plane sections of the fractal structures induced by the competition between decoherence and purification. We want now make a 3D representation based on the embedding $\mathrm{p}(\mathbb H) \to \mathbb R^3$ defined by the coordinates:
\begin{eqnarray}
  X(\zeta) & = & \RE(\zeta) = \RE(z) \cos \lambda \\
  Y(\zeta) & = & \IM_1(\zeta) = \IM(z) \cos \lambda \\
  Z(\zeta) & = & -|\zeta-\CO(\zeta)| = -|z| \sin \lambda
\end{eqnarray}
with $\zeta = z e^{\jquat \lambda}$ ($z \in \mathbb C$) (the spherical coordinates being $(|z|,\arg z,\lambda)$). Quaternionic fractal sets corresponding to the pure dephasing and to the map (\ref{du}) are represented fig. \ref{Mandelbulb}.
If usual Mandelbulbs present fractal protuberances, these ones present fractal alveoli. Maybe these structures should be called ``Mandelcheeses''.\\

\begin{figure}
  \includegraphics[width=8cm]{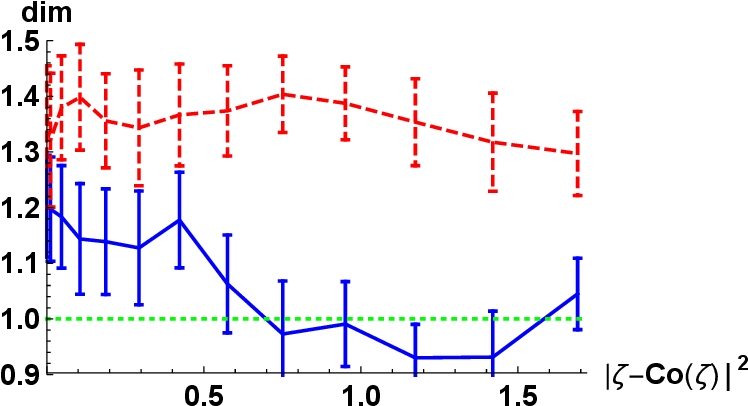}
  \caption{\label{dimH} Estimation of the upper-box-counting dimensions of the sections $|\zeta-\CO(\zeta)|=C^{ste}$ of the Mandelbulb like borders (blue plain line for the pure dephasing process and red dashed line for the decoherence process eq. \ref{du}). A dimension equal to 1 corresponds to a border being a simple curve whereas a non integer value of the dimension corresponds to a fractal border. Note that due to the difficulty to make a precise numerical estimation of a fractal dimension, the values appearing in these graphs are rough estimates but the variations are meaningful.}
\end{figure}

The fractality seems evolve with $|\zeta-\CO(\zeta)|$ as shown fig. \ref{dimH}.
In contrast with the case of the decoherence process eq. \ref{du}, for the case of the pure dephasing process we see after an initial plateau that the fractality decreases with growing values of $|\zeta_0-\CO(\zeta_0)|^2$ (the concurrence of the initial equivalent entanglement). For a square concurrence larger than $0.8$, the border seems to be a simple curve (as also shown fig. \ref{purityMandeldephasing}).

\section{Conclusion}
The competition between decoherence processes and purification protocols on a qubit can be represented by nonlinear maps onto the quaternion space $\mathbb H$. These maps belong to the Julia map family. The border between the area dominated by the purification and the area dominated by the decoherence are like Mandelbulbs. Due to this fractal structure, it is difficult to know if an initial state will be at the end purified or mixed by the competition between the two processes. This is particularly the case for states in the neighbourhood of the pure state space which is a highly fractalised region. In this paper we have considered that the evolution operator $U$ is still the same at each iteration. In applications to quantum computation and quantum control, the Hamiltonian is time-dependent and the evolution operator changes at each iteration. Moreover, we can also modify at each iteration the purification protocol to help the control (by varying the parameters $\alpha$, $\beta$, $\gamma$ and $q$, or by changing the basis of purification (which is always $(|0\rangle,|1\rangle)$ in this paper)). The behaviour of the competition will be more complicated but maybe this could be help to solve quantum control problems in presence of decoherence processes.

\section*{Acknowledgments}
The author acknowledge support from ISITE Bourgogne-Franche-Comt\'e (contract ANR-15-IDEX-0003) under grants from I-QUINS and GNETWORKS projects, and support from the R\'egion Bourgogne-Franche-Comt\'e under grants from the APEX project. Numerical computations have been executed on computers of the Utinam Institute supported by the R\'egion Bourgogne-Franche-Comt\'e and the Institut des Sciences de l'Univers (INSU).

\end{document}